# Exposing and extending the interior waves field by transformation materials

**Jiangchao Shi[1,2], Xiaobo Yang[1,2] and Jin Hu[1,2,*]**

1 Beijing Institute of Technology, School of Information and Electronics, 100081, Beijing, China
2 China Beijing Key Laboratory of Fractional Signals and Systems, 100081, Beijing, China

* Corresponding author E-mail: bithj@bit.edu.cn

## Abstract

Based on transformation optics, a strategy is proposed to expose the inner one-dimensional space of a wave field inside a beam volume to the surface of the propagation medium and extend the space from one-dimensional to two-dimensional, allowing the corresponding field distribution to be detected directly and more subtly, which is important in optical signal processing. The method is applied to the quadratic graded index lens to construct a new graded index lens, and its enhanced chirpyness detection ability is demonstrated by numerical simulation.

## 1. Introduction

The optical signal processing deals with the information carried by the electromagnetic beam volume [1]. In some cases, the interested wave behaviours are located in the inner parts of the beam and distributed along the propagation path. For example, in the graded index (GRIN) lens–based chirpyness detection [2-4], the most important information is the impulses' positions at the optical axis, which locates inside the three-dimensional (3D) lens body. Another example of inner information of waves is the spatially nonuniform polarization or helical phase inside the vortex beam [5, 6], where the corresponding profile of the physical statuses along the beam propagation path is significant in identifying or describing the beam. In these circumstances, the direct and stable exploration of the inner information becomes inconvenient or impossible; because the detection probe cannot be configured inside the propagation media, as well as arranged along the propagation route, without disturbing the wave field. Some auxiliary methods, such as semi-lens or tomography technology, are suggested to assist in the inner detections [3, 7], but with the cost of lower precision and inconvenience.

Hence, an interesting and useful question arises: Can we directly monitor the wave behaviours inside the beam volume? In this study, we proposed a strategy based on transformation optics (TO) [8, 9] to bare and enlarge an internal part of a beam to the surface of the propagation media, making the direct observation and measurement of the corresponding wave field possible. The idea is based on the ability of TO to manipulate the electromagnetic wave behaviours in a prescribed pattern by mapping the propagation space from one to another. With a suitable space mapping, a line segment inside the original propagation medium is transferred to the surface of the medium and is extended to be a plane; according to TO principle, the corresponding wave field originally distributed in the line is also shifted to the surface plane; thus, the design goal is achieved. As an application example of this methodology, the conventional quadratic GRIN lens is transformed to a new GRIN lens and is used for the detection of the chirpyness of an input chirp signal with improved convenience and precision. The proposed idea may open a new way for the manipulation and detection of unusual optical signals.

The rest of the paper is organized as follows. In Section 2, the space mapping adopted in this TO-based design strategy is presented, which can be used for any wave fields; then as an example, in Section 3 the strategy is applied to the GRIN lens for chirpyness detection, and the enhanced abilities of the designed lens are demonstrated by numerical simulations. Finally, the discussions and conclusions are provided in Section 4.

**2. The space mapping of the exposure**

We start our design strategy by considering the appropriate space mapping used for this particular aim in the framework of TO. For clarity, the two-dimensional (2D) mapping is analyzed, and the 3D medium can be obtained by rotating or stacking the 2D transformation materials. According to TO theory, the isotropic transformation media need the space mappings to be conformal if the original material is isotropic. To this end, the function of complex variable is utilized, and the conformal transformation we adopted is [10]

$$w = \sin(z) \tag{1}$$

where $w = u + vi$ and $z = x + yi$ are the points in original and transformed complex planes, respectively. It can be shown that this transformation splits the radial from $(-1,0)$ to $(-\infty, 0)$ to two and rotates them around point $(-1,0)$ in $w$ space to form a new line in $z$ space as $y = -\pi/2$, which is a border of the area, as shown in figure 1. That is, part of the inner line that hides inside the $w$ space is exposed to face outside in the $z$ space. According to TO principle, the wave field distributed in the original inner line will now be point-to-point mapped to the corresponding border and thus can be measured or observed directly without disturbing the field.

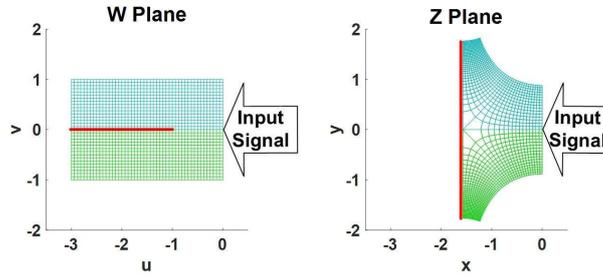

**Figure 1.** Schematic illustration of the conformal transformations of the exposed inner area. The red line inside the $w$ space is mapped to be the red border in $z$ space. Both right borders are kept straight, which can be set as the input facet.

Furthermore, if the transforming plane is the meridian plane, and the original line lies on the optical axis, the corresponding 3D case is equivalent to rotate the planes around the optical axis, and then the inner line is not only exposed to the surface, but also extended from one-dimensional (1D) space to 2D space, as shown in figure 2. This dimensionality extension implies that the information carried by the original wave field can be revealed in a much more exquisite manner.

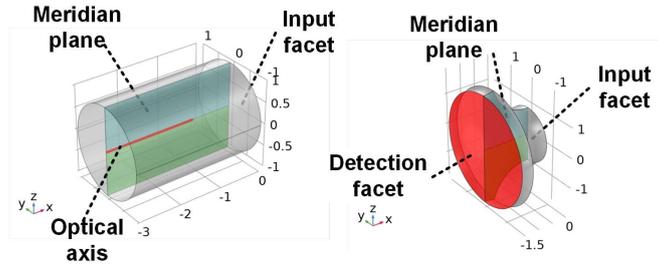

**Figure 2**. The 3D space mapping corresponding to figure 1. Part of the inner line in $w$ space is extended from 1D to 2D and becomes a surface in $z$ space.

Before exploring its application, we first investigate some properties of the proposed transformation medium. The isotropic principal stretch of the space deformation is $\lambda = |dz/dw| = 1/|\cos(z)|$, and the refractive index of the transformation medium is $n = n_0/\lambda$, where $n_0$ is the refractive index of the original medium [11]. The principal stretch is a very important quantity in the context of conformal TO, which can completely describe the properties of the transformations of the medium and physical fields.

It is noted that the right borders in the deformed plane in figure 1 are maintained to be in a straight line, which means that the input facet can be kept as a flat plane, as shown in figure 2, which may have more advantages in practice compared to those transformation media with curve input facets [2-4, 12]. The $\lambda$ distribution on the input facet in the transformation space is shown in figure 3(a). In the paraxial applications, the stretches are approximately equal to 1, which implies that the input signal can be considered as not influenced by the input facet deformation. In contrast, the $\lambda$ distribution on the detection facet, which is transformed from the inner line, varies greatly in the paraxial area as shown in figure 3(b). There exists an area in the detection facet where $\lambda$ is larger than 1, which indicates that the information there is magnified along radial direction, which can help improve the detection precision.

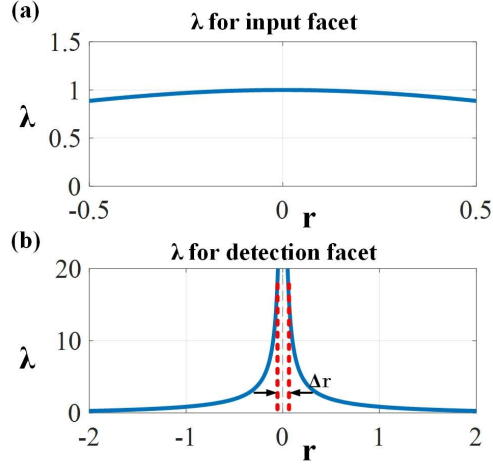

**Figure 3.** The principal stretch of the (a) input and (b) detection facets. In (b), the dotted lines show that the $\lambda$ changes sharply in a small area, and it can be truncated to remove the singularity.

The singularity of infinite $\lambda$ or zero refractive index at the centre of the detection facet can be removed without significant impact on the wave field, because as shown in figure 3(b), the curve becomes very sharp near the centre, which means that the extreme refractive index is centralized in a very small area around the facet centre. When replacing the ideal refractive index with a reasonable value in this small area, its tiny size will ensure the low effect on the wave propagation.

## 3. An example application of the strategy in designing GRIN lens for chirpyness detection

### 3.1 Chirpyness detection principle based on GRIN lens

The proposed design strategy is a general one and can be used in many occasions to enhance the detection of information carried by the wave field originally distributed in the interior zone of a beam. Here, we apply it to a conventional GRIN lens and derive a new kind of GRIN lens that can detect the chirpyness of an input chirp signal with a much higher precision and new capability on the surface of the lens.

Chirp signals have some important applications in signal processing. In manoeuvring target detection and imaging in the fields of radar, sonar, etc, the velocity and acceleration information of the target can be obtained by modelling the radar echoes as chirp signals [13-15], since this information corresponds to the chirpyness of chirp signals. The Newton's ring can also be modelled as chirp and its chirpyness is an important parameter in related optical measurements [16].

Most chirpyness estimating methods are based on time-frequency representations such as Wigner-Ville distribution (WVD) [17] and fractional Fourier transform (FrFT) [18], higher-order ambiguity function (HAF) [19], higher-order phase function [20], nonlinear least squares method (NLSE) [21], and so on. These algorithms are time consuming, because they must process 2D searches or other exhaustive searches even for the 1D problems. Alternatively, the GRIN lens based chirpyness detection provides a fast and convenient optical approach in this field, is worth further exploring.

A quadratic GRIN lens has a refractive index distribution as $n^2(r) = n_1^2[1 - (n_2/n_1)^2 r^2]$ [22], where $r$ represents the radial distance from the optical axis, and $n_1$, $n_2$ are the medium parameters. The chirpyness detection ability of this lens is based on its FrFT operation inside the lens body. As a generalization of the classic Fourier transform (FT), FrFT plays an important role in optical information processing [23] and can be expressed as (take the 1D case as an example)

$$g_p(u) = F^p[f(x)](u) = \int_{-\infty}^{\infty} f(x) K_p(x,u) dx, \qquad (2)$$

and the inverse FrFT is

$$f(x) = (F^{-1})^p[g_p(u)](x) = \int_{-\infty}^{\infty} g_p(u) K_{-p}(x,u) du, \qquad (3)$$

where $p$ is a real number called the order of the FrFT, and the kernel is given by

$$K_p(x,u) = A_{p\lambda_0} e^{-\frac{2\pi}{s^2}ixu-csc\alpha} e^{\frac{i\pi}{s^2}(x^2+u^2)cot\alpha}, \qquad (4)$$

where $A_{p\lambda_0} = 1/(is\sqrt{2\pi sin\alpha})$ with $\alpha = p(\pi/2)$, and $s^2 = \lambda_0\sqrt{1/n_1 n_2}$ is the scale factor of the coordinates, with $\lambda_0$ being the wavelength. If $p = 1$ or $\alpha = \pi/2$, FrFT is reduced to the FT.

The inverse FrFT indicates that any signal can be decomposed in terms of the orthonormal chirp bases $K_p$; thus, if the signal itself is a chirp,

$$f(x) = e^{i\frac{1}{2}mx^2+im_0 x}, \qquad (5)$$

where $m$ is the chirpyness, and $m_0$ is a constant, its FrFT domain $g_p(u)$ will be an impulse if the order $p$ making the base has the same chirpyness. This is exactly similar to a sinusoidal signal that will become an impulse in its FT result. Setting the chirpyness of the kernel equals that of the input chirp signal, equation (4) and (5) can give

$$m = -\frac{2\pi}{s^2}cot\alpha = -\frac{2\pi}{s^2}cot\left(p\frac{\pi}{2}\right). \qquad (6)$$

This relation shows that if the corresponding FrFT order can be found, then the input chirpyness will be derived. For the quadratic GRIN lens, the input signal's FrFT with order $p(0 \leq p \leq 1)$ can be detected at the facet with distance to the input facet as [22]

$$l = pL, \qquad (7)$$

where $L = (\pi\sqrt{n_1/n_2})/2$ is the focal length of the lens, as shown in figure 4. Specifically, for an input chirp signal, if we can find the impulse location in the lens body, then the corresponding FrFT order can be obtained from equation (7), and in turn the chirpyness of the input chirp can be derived from equation (6).

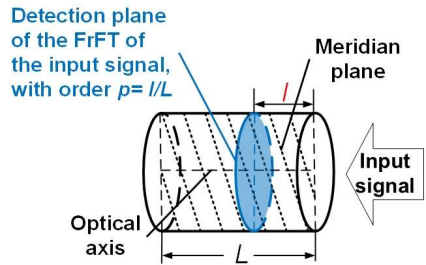

**Figure 4.** The schematic of determining optical FrFT with different orders in a quadratic GRIN lens.

In practice, especially in 3D cases, however, obtaining the impulse location inside the GRIN lens body is not an easy task. One way to disclose the inner part of the lens is to cut off the lens along one meridian plane to constitute a semi-lens to expose the meridian, which will lose the completeness of the lens and lead to a lower detection precision. The tomography technology can be used to detect the inner information facet-by-facet along the optical axis; however, the exhaustive searching is quite time consuming.

To overcome these defects, the proposed design strategy is applied to the conventional quadratic GRIN lens. As mentioned previously, an area in the detection surface of the designed lens is amplified, which can improve the detection accuracy when the impulse positions fall into the area.

3.2 Numerical Simulations of the proposed lens

The proposed lens is validated by numerical simulations based on COMSOL Multiphysics. For the conventional GRIN lens, we set $n_1 = 6$, $n_2 = 1.6449$, and the normalized focus length $L = 3$. The refractive index distributions of the original and new lens are shown in figure 5. The simulation results, with normalized wavelength of 0.2 and 0.6 for 2D and 3D cases, are shown in figures 6 and 7 and tables 1 and 2, respectively. The impulses' locations are exposed from the inner optical axis in the conventional lens to the surface in the proposed lens and can be used to obtain the input chirpyness according to the mapping relation. The proposed lens can have multiple detection lines in the detection plane, which also helps reduce the errors. It is verified that the proposed lens can indeed expose and dimensionality extend the inner information, and the chirpyness detection precision is improved.

The input chirp signals are required to be centred in the input facet to make the impulses located at the optical axis. In case there exists a small offset error in the input chirp signal, the detection in the 1D optical axis is impossible to be aware of the offset and compensate for the error. The proposed GRIN lens, however, is capable of alerting the centre shift of the input. This is because the optical axis is extended to the 2D space, and the nonuniform distribution of the wave field surrounding the axis is mapped to the plane, which directly reveals the asymmetry of the input signal. The simulation results of such samples are shown in figure 8. Because of the offset, the locations of the peak value in the detection facet are no longer symmetrical and are directly related to the offset directions. The asymmetrical peak locations can also be used to compensate the error caused by the input offset, as shown in table 3(the corresponding errors in the conventional GRIN lens with nonzero input offsets are greater than 20%).

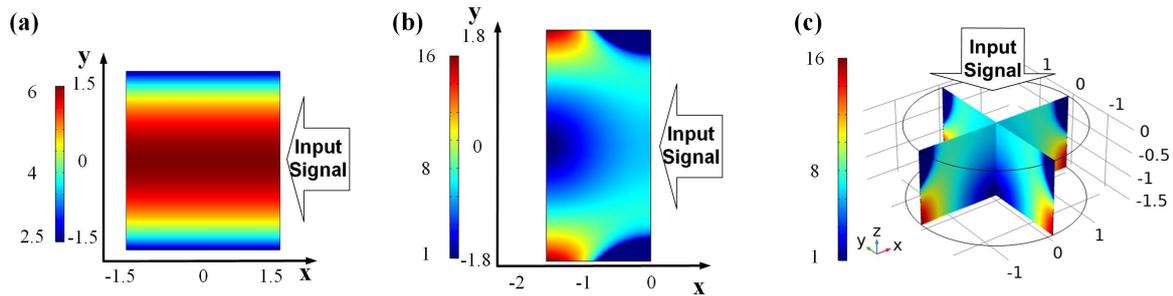

**Figure 5.** The refractive index distribution in the meridian planes of (a) conventional GRIN lens and (b) the proposed lens. (c) is the 3D view of the refractive index distribution of the proposed lens. The proposed lens is complemented to be a regular shape for convenience.

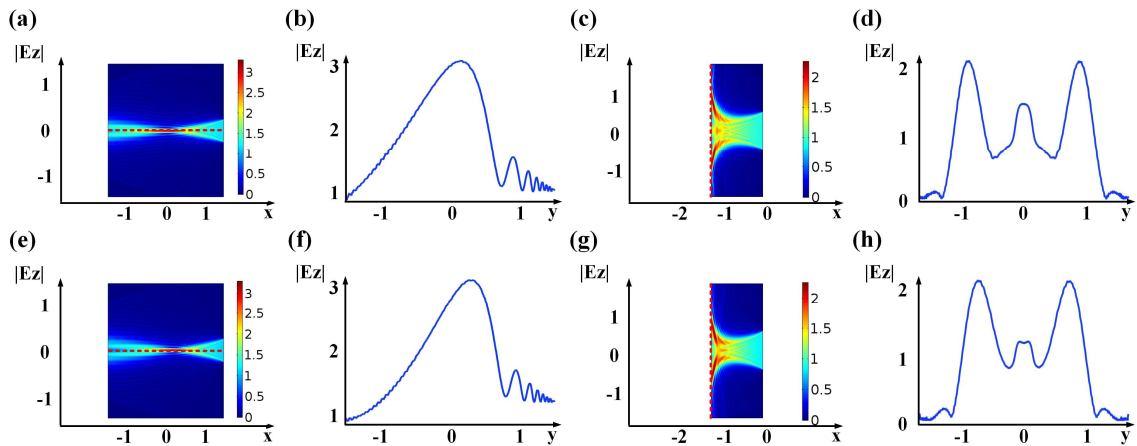

**Figure 6.** The 2D simulation results of the wave fields and corresponding amplitude distributions along the detection lines (dotted lines). (a, b) conventional lens with chirpyness $m = 100$; (c, d) proposed lens with chirpyness $m = 100$; (e, f) conventional lens with chirpyness $m = 120$; (g, h) proposed lens with chirpyness $m = 120$.

Table 1. The 2D simulation results of the conventional lens and the proposed lens.

| Lens | Input test chirpyness | Obtained chirpyness | Chirpyness error(%) |
|---|---|---|---|
| Conventional | 100 | 116.5710 | 16.57 |
| Proposed | | 103.0056 | 3.58 |
| Conventional | 120 | 131.0944 | 9.25 |
| Proposed | | 120.4041 | 0.26 |

Table 2. The 3D simulation results of the conventional lens and the proposed lens.

| Lens | Input test chirpyness | Obtained chirpyness | Chirpyness error (%) |
|---|---|---|---|
| Conventional | 30 | 41.0405 | 36.80 |
| Proposed | | 29.5546 | -1.48 |
| Conventional | 40 | 51.9276 | 29.82 |
| Proposed | | 40.0902 | 0.23 |

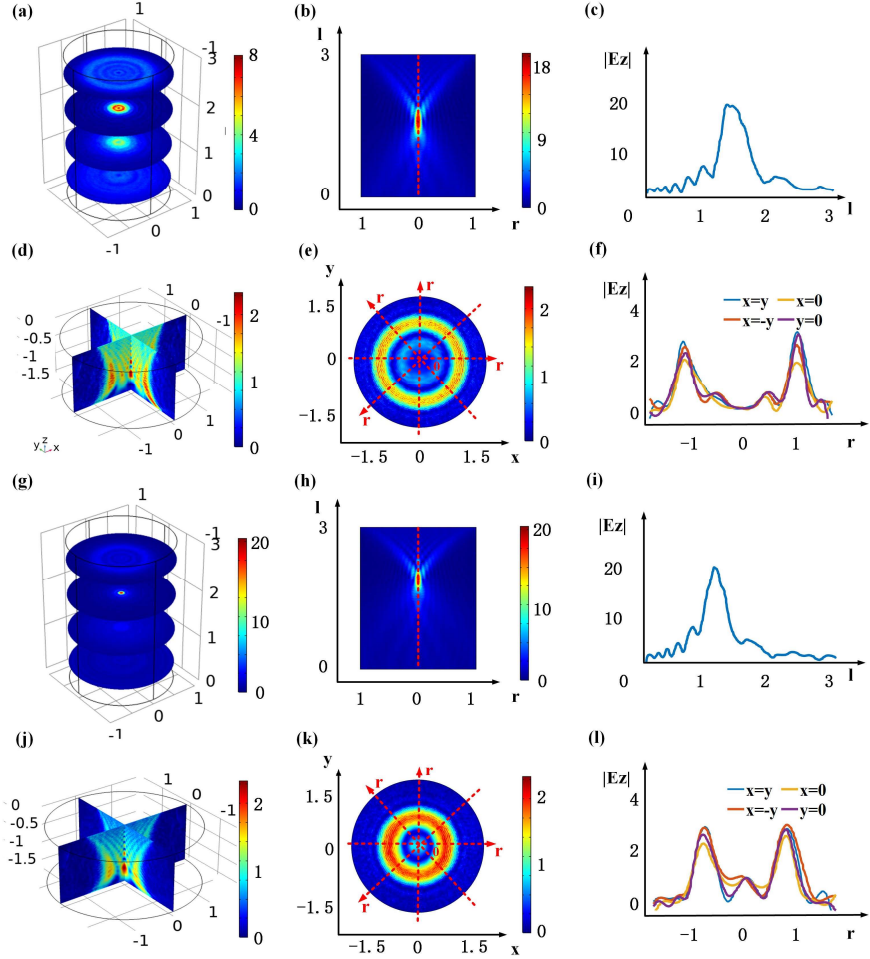

**Figure 7.** The 3D simulation results of the lenses. The left column is the 3D view of the wave field distribution, the middle column is the wave field distribution in the detection plane, and the right column is the envelope of corresponding amplitude distributions along the detection lines (dotted lines). (a-c) conventional lens with chirpyness $m = 30$; (d-f) proposed lens with chirpyness $m = 30$; (g-i) conventional lens with chirpyness $m = 40$; (j-l) proposed lens with chirpyness $m = 40$.

Table 3. The simulation results of the input signal with offsets in the proposed lens.

| Input offset | Input test chirpyness | Obtained chirpyness (high/low) (average) | Chirpyness error (%) (high/low) (average) |
|---|---|---|---|
| (0, 0) | 30 | 30.2293/29.6247<br>29.9270 | 0.76/-1.25<br>-0.24 |
| (0.1, 0) | | 27.7423/38.1407<br>32.9415 | -7.53/27.14<br>9.81 |
| (0.1, 0.1) | | 26.5956/36.4636<br>31.5296 | -11.35/21.55<br>5.10 |

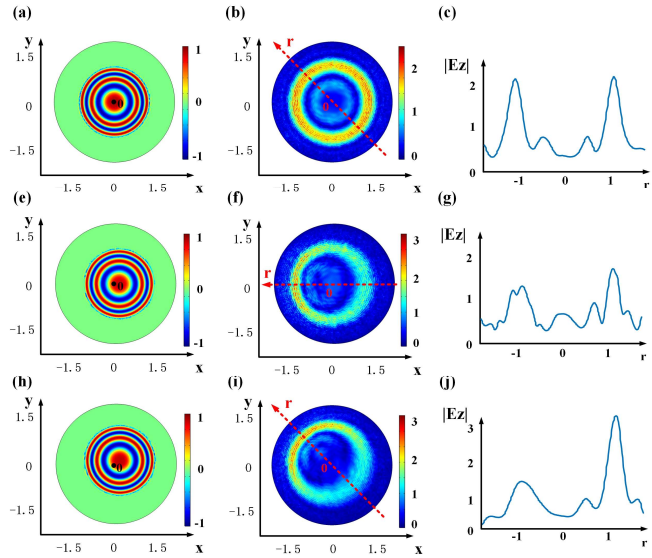

**Figure 8.** Simulation of input offsets in the proposed lens. The left column is the input facet, the middle column is the detection plane, and the right column is the envelope of corresponding amplitude distributions along the detection lines (dotted lines, which can be found as the symmetry axis of the amplitude distribution there). Note that the left and middle columns are viewed in opposite directions. (a-c) no offset; (d-f) offset centre is (0.1, 0); (g-i) offset centre is (0.1, 0.1).

## 4. Discussion and Conclusion

In conclusion, with the help of TO, we have proposed a strategy to expose the inner 1D space of a wave field inside a beam volume to the surface of the medium and extend the exposed space dimensionality from 1D to 2D, allowing the corresponding field distribution to be detected directly. This transference is useful in optical signal processing, where in many cases the information is carried by the inside wave behaviours and is inconvenient or impossible for direct detection. The proposed strategy is applied to the quadratic GRIN lens to construct a new GRIN lens, and its enhanced chirpyness detection ability is demonstrated, including the on-surface direct detection, higher detection precision, and input offset perceptibility. As the chirpyness detection is an important issue in optical signal processing, it is expected that the proposed lens can offer a fast and easy tool in this field.

The proposed GRIN lens is isotropic and therefore has advantages in broadband and low loss practical applications, and the fabrication of it is possible. For example, the 3D printing technology allows structures to be fabricated directly from digital models without the need for moulds, making them ideal for components with high geometric or index complexity. This technology has been applied to the manufacture of metamaterials and artificial electromagnetic (EM) medium[24-26]. The GRIN lens manufacturing at the millimetre level has now been realized[27, 28].

It is worth emphasizing that the design strategy itself is a general frame in exposing, transferring, or extending particular parts of a wave field. It is possible to design more novel devices by applying the idea to other GRIN media or introduce other similar space mappings to enrich the strategy. The related investigation is in progress.

## Acknowledgment

This work was supported by the National Natural Science Foundation of China (61575022, 61421001, 61975015).